\documentclass[a4paper,11pt]{article}
\usepackage{pos}
\usepackage{wrapfig}

\title{Recent results from DANSS}

\author*[a,b,1]{Igor Alekseev}

\affiliation[a]{P.N. Lebedev Physical Institute of the Russian Academy of Sciences,\\
  Leninsky prospect 53, Moscow, Russia}

\affiliation[b]{NRC "Kurchatov Institute",\\
Kurchatov square 1, Moscow, Russia}

\note{For the DANSS collaboration}

\emailAdd{igor.alekseev@itep.ru}

\abstract{DANSS is a solid state scintillator neutrino spectrometer placed at
a small distance from the commercial nuclear reactor of Kalininskaya
NPP. The distance from the detector to the center of the reactor core
can be changed online in the range 10.9-12.9 m. This fact together with
a very high neutrino counting rate (more than 5000 events per day) and
low background makes DANSS an ideal detector to search for neutrino
oscillations in 1~eV$^2 \Delta m^2$ range. 

We report the results based on the statistics of 6 million events,
obtained between April 2016 and March 2022. The results include limits in
the short range oscillation parameter space, fuel evolution studies
and the bump in the neutrino spectrum. The talk will also cover our
plans of the detector upgrade.}

\FullConference{%
  Neutrino Oscillation Workshop-NOW2022\\
  4-11 September, 2022\\
  Rosa Marina (Ostuni, Italy)}

\begin{document}
\maketitle

\begin{figure}[h]
\vspace{-0.2cm}
\begin{center}
\includegraphics[width=\textwidth]{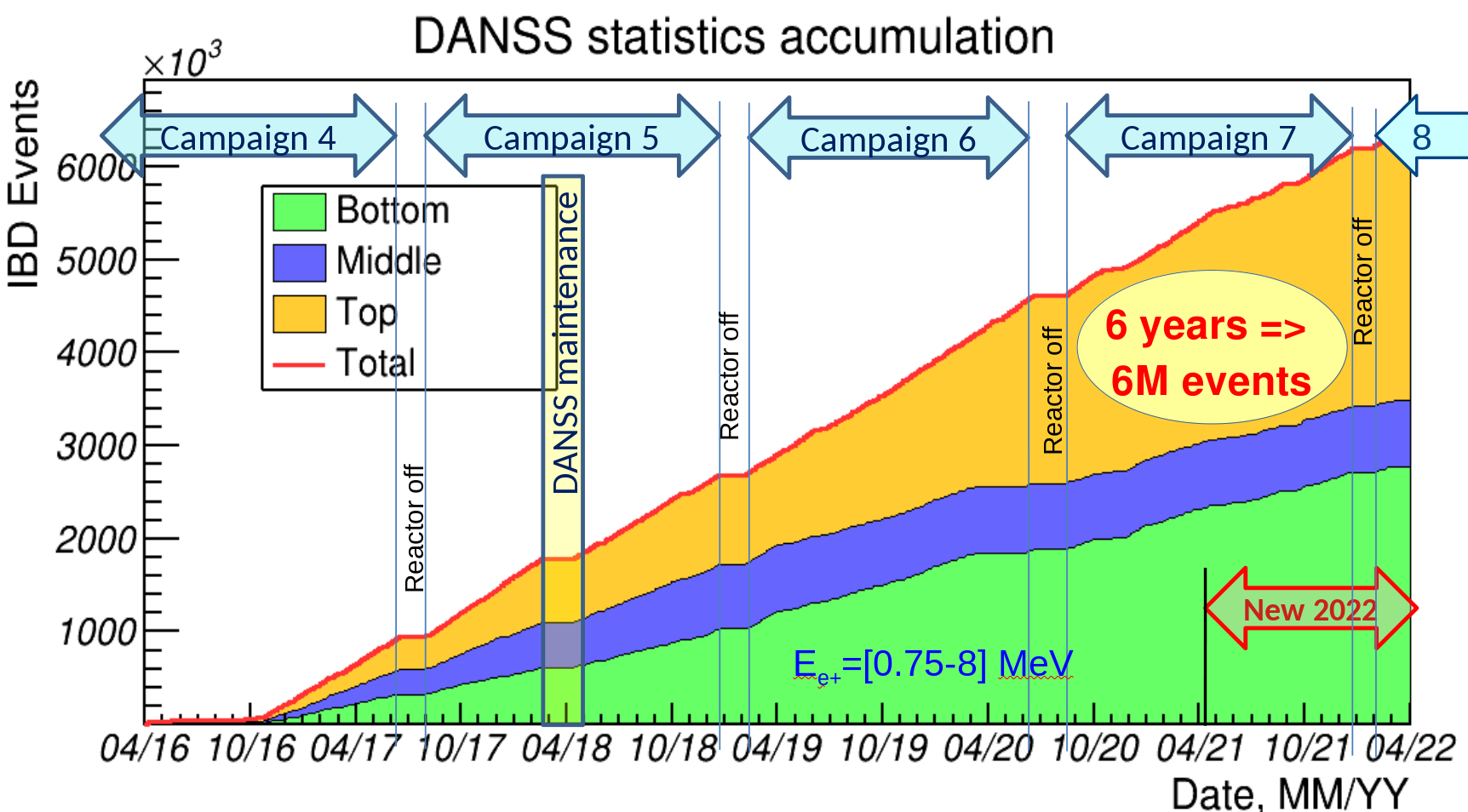}
\end{center}
\vspace{-0.2cm}
\caption{\label{fig:stat}IBD statistics accumulation during 6 years of DANSS operation}
\vspace{-0.2cm}
\end{figure}
\begin{figure}[h]
\begin{center}
\vspace{-0.2cm}
\includegraphics[width=0.41\textwidth]{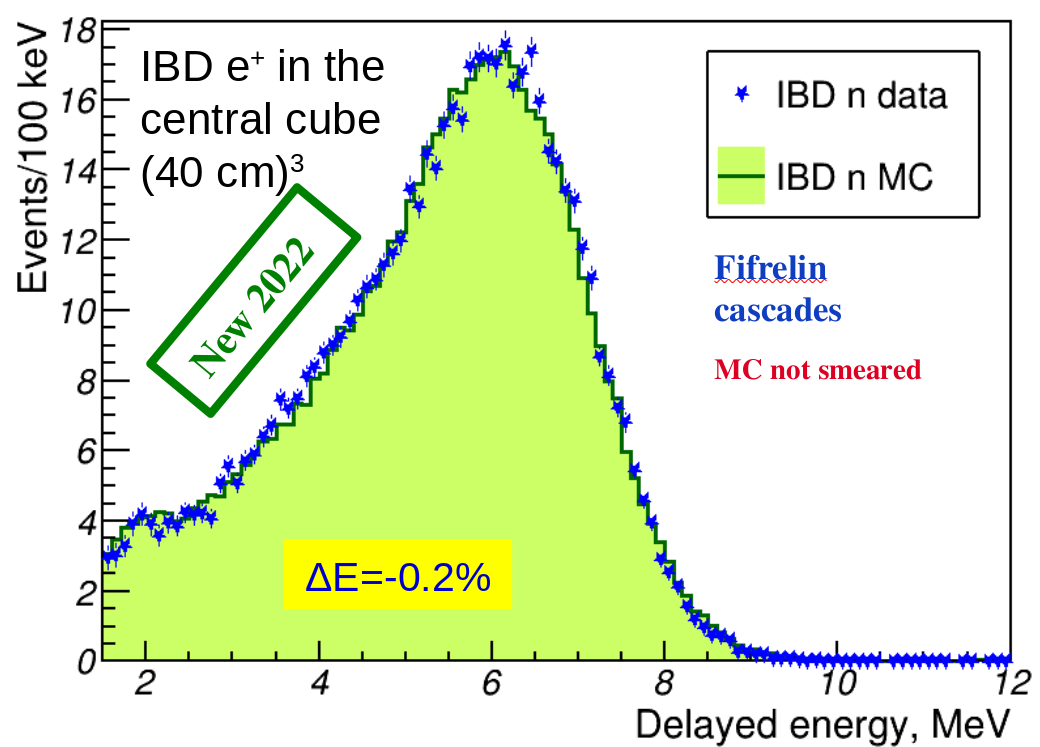}%
\hspace{0.01\textwidth}%
\includegraphics[width=0.25\textwidth]{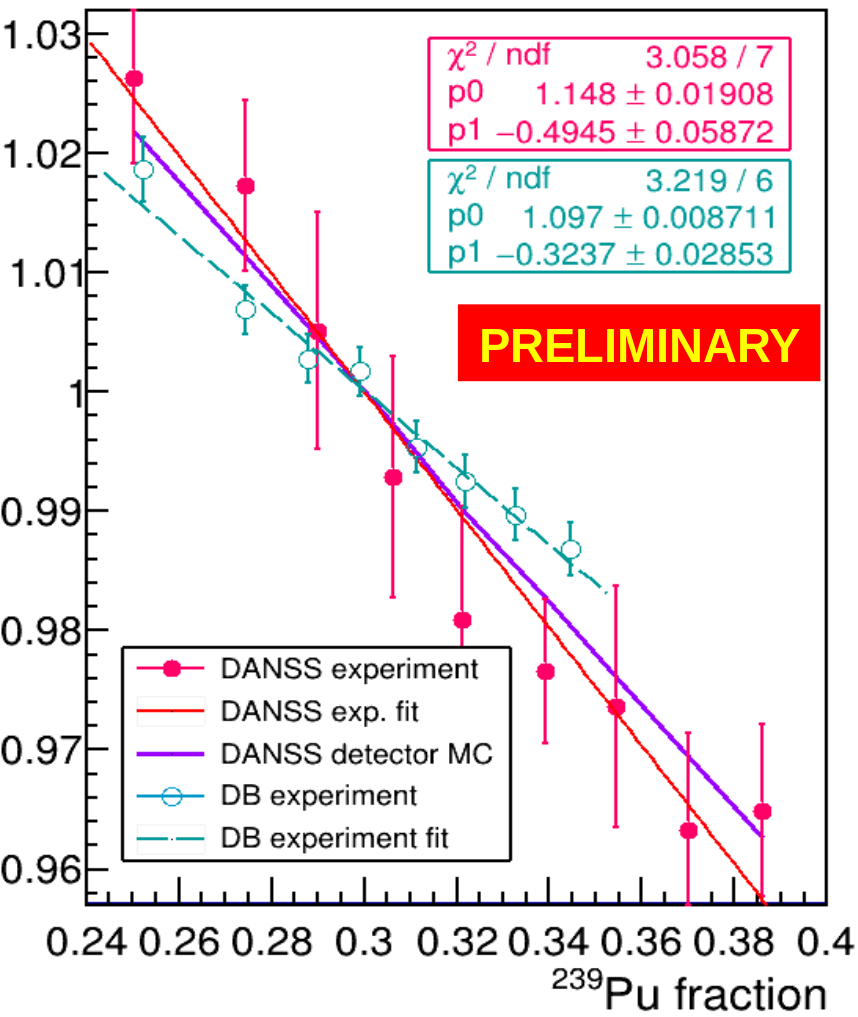}%
\includegraphics[width=0.32\textwidth]{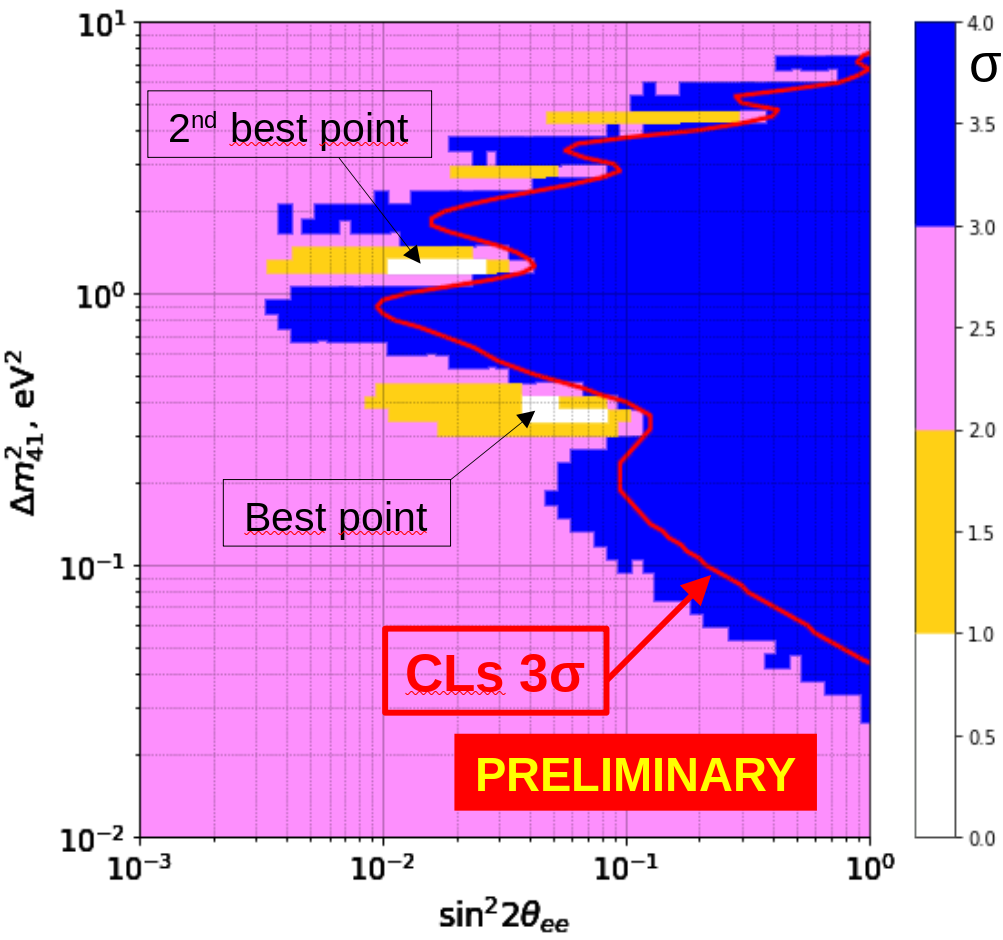}
\vspace{-0.2cm}
\end{center}
\caption{Spectrum of neutron capture for IBD events with positron vertex in the central (40cm)$^3$ cube (left).
Neutrino counting rate dependence over fission fraction of $^{239}$Pu normalized to the counting rate at 
30\% fraction (middle). Feldman and Cousins analysis of DANSS data (right).}
\label{fig:IBD}
\end{figure}

Details of the DANSS detector and physics results of the first year of its operation could be
found elsewhere  \cite{DANSS_JINST, DANSS_PLB} as well as the results obtained by the previous year
\cite{DANSS_TAUP21, DANSS_NUFACT21}. This short contribution is concentrated over
our progress during the last year. Our progress in the inverse beta-decay (IBD) statistics accumulation is illustrated in
fig.~\ref{fig:stat}. Now we have data for 3 full fuel campaigns and 4 reactor off periods. An important progress is
reached in the understanding of our calibration. In addition to the reactions already used for calibration purposes delayed
event spectrum is also analyzed as a calibration source (fig.~\ref{fig:IBD} left). Now the calibration set includes $^{12}$B decays
from two reactions, induced by atmosphere muons, $n+^{12}$C$\to^{12}$B$+p$ and $\mu^-$ capture by $^{12}$C; $^{22}$Na and $^{60}$Co 
radioactive sources; neutrons from $^{248}$Cm fission and IBD, stopped muons decays. $^{12}$B decay data is used to set the scale, 
because behavior of the produced electron is the most similar to IBD positron we need to measure and this data 
is accumulated uniformly during the run. No additional smearing is added in the Monte-Carlo simulations any more for a good 
reproduction of the experimental data. The scale of all the sources in the calibration set agrees within $\pm 0.2$ with an 
exception of $^{22}$Na, which has an offset 1.8\%. The problem could be in contamination of the sample with $^{26}$Al with 
slightly different energy of the decay. We keep systematic error 2\% in the energy scale until we find a solution of this problem.

Continuous reactor monitoring during 3 full fuel campaigns allows us to make a study of counting rate and 
neutrino spectrum evolution with the change in the fuel composition. The rate dependence over fission fraction of
$^{239}$Pu is shown in fig.~\ref{fig:IBD} (middle). Our data demonstrate slope slightly steeper than the slope coming from
MC-simulations with HM model \cite{Huber, Mueller}, while Daya Bay data show less steep slope \cite{DB_fuel}. We also have new 
results in light sterile neutrino search. After the new data was included into the analysis the 90\% confidence level
limit to sterile neutrino parameters in the region of $\Delta m^2 \sim 0.9$~eV$^2$ became as stringent as 
$\sin^2 2\theta < 0.004$ (Gaussian CLs method), but two points with close $\Delta \chi^2 \sim -10$ manifested themselves. 
A Feldman and Cousins method was used to obtain their significance. The result is shown in fig.~\ref{fig:IBD} (right). 
A dark blue area corresponds to $3\sigma$-limit. Much more conservative $3\sigma$-limit from Gaussian CLs method
is shown by red line for comparison. The best fit point has significance $2.35 \sigma$, which is much less than 
we need to claim an indication of the 4th neutrino existence. 

The collaboration appreciates the permanent assistance of the KNPP administration and Radiation Safety Department staff. 
This work is supported by Ministry of Science and Higher Education of the Russian Federation under Contract No. 075-15-2020-778.


\begin{thebibliography}{9}
\bibitem{DANSS_JINST} I. Alekseev et al., \emph{DANSS: Detector of the reactor AntiNeutrino based on Solid Scintillator}, 
	\href{https://doi.org/10.1088/1748-0221/11/11/P11011}{{\it JINST} {\bf 11} (2016) P11011} [{\tt physics.ins-det/1606.02896}]
\bibitem{DANSS_PLB} I. Alekseev et al., \emph{Search for sterile neutrinos at the DANSS experiment}, 
	\href{https://doi.org/10.1016/j.physletb.2018.10.038}{{\it Phys. Lett. B} {\bf 787} (2018) 56} [{\tt hep-ex/1804.04046}]
\bibitem{DANSS_TAUP21} Igor Alekseev for the DANSS Collaboration, \emph{Antineutrino spectrometer DANSS — 5 years of running},
	\href{https://doi.org/10.1088/1742-6596/2156/1/012100}{\emph{Journal of Physics: Conference Series} {\bf 2156} (2021) 012100}
\bibitem{DANSS_NUFACT21} I.G. Alekseev and N. Skrobova, \emph{Recent results of the DANSS experiment},
	\href{https://doi.org/10.22323/1.402.0143}{\emph{PoS} {\bf 402} NuFact2021 (2022) 143}
\bibitem{Huber} P. Huber, \emph{Determination of antineutrino spectra from nuclear reactors}, 
	\href{https://doi.org/10.1103/PhysRevC.84.024617}{\emph{Phys.Rev. C} \bf{84} (2011) 024617} [{\tt hep-ph/1106.0687}]
\bibitem{Mueller} Th.A. Mueller et al., \emph{Improved predictions of reactor antineutrino spectra}, 
	\href{https://doi.org/10.1103/PhysRevC.83.054615}{\emph{Phys.Rev. C} \bf{83} (2011) 054615} [{\tt hep-ex/1101.2663}]
\bibitem{DB_fuel} F. P. An et al., \emph{Evolution of the Reactor Antineutrino Flux and Spectrum at Daya Bay}, 
	\href{https://doi.org/10.1103/PhysRevLett.118.251801}{\emph{Phys. Rev. Lett.} \bf{118} (2017) 251801} [{\tt hep-ex/1704.01082}]
\end{thebibliography}
\end{document}